\documentclass[aps,prl,reprint,superscriptaddress]{revtex4-2}

\usepackage[english]{babel}
\usepackage{graphics}
\usepackage{array}
\usepackage{graphicx}
\usepackage{booktabs}
\usepackage{todonotes}
\usepackage{amsfonts,amssymb,amsmath}
\usepackage{color}
\usepackage{threeparttable}
\usepackage{physics}
\usepackage{comment}
\usepackage[separate-uncertainty,per-mode=fraction]{siunitx}

\definecolor{darkblue}{rgb}{0.0,0.0,0.4}
\definecolor{darkgreen}{rgb}{0.0,0.4,0.0}
\definecolor{darkred}{rgb}{0.6,0.0,0.0}
\usepackage[bookmarksopen]{hyperref}
\hypersetup{colorlinks,linkcolor=darkblue,citecolor=darkblue,urlcolor=darkblue}

\newcommand{\gs}{$\mathrm{X}^2\Sigma^+$}
\newcommand{\exs}{$\mathrm{A}^2\Pi_{1/2}$}
\newcommand{\baff}[1]{${}^{#1}\mathrm{BaF}$}

\newcolumntype{M}[1]{>{\centering\arraybackslash}m{#1}}

\makeatletter
\newcommand\thefontsize[1]{{#1 The current font size is: \f@size pt\par}}
\makeatother

\begin{document}

\title{Molecular laser cooling using serrodynes:\\Implementation, characterization and prospects}

\author{Felix Kogel}
\thanks{These two authors contributed equally}
\affiliation{5. Physikalisches  Institut  and  Center  for  Integrated  Quantum  Science  and  Technology, Universit\"at  Stuttgart,  Pfaffenwaldring  57,  70569  Stuttgart,  Germany}

\author{Tatsam Garg}
\thanks{These two authors contributed equally}
\affiliation{5. Physikalisches  Institut  and  Center  for  Integrated  Quantum  Science  and  Technology, Universit\"at  Stuttgart,  Pfaffenwaldring  57,  70569  Stuttgart,  Germany}

\affiliation{Vienna Center for Quantum Science and Technology, Atominstitut, TU Wien,  Stadionallee 2,  A-1020 Vienna,  Austria}

\author{Marian Rockenh\"auser}
\affiliation{5. Physikalisches  Institut  and  Center  for  Integrated  Quantum  Science  and  Technology, Universit\"at  Stuttgart,  Pfaffenwaldring  57,  70569  Stuttgart,  Germany}

\affiliation{Vienna Center for Quantum Science and Technology, Atominstitut, TU Wien,  Stadionallee 2,  A-1020 Vienna,  Austria}

\author{Sebasti\'an A. Morales-Ram\'irez}
\affiliation{5. Physikalisches  Institut  and  Center  for  Integrated  Quantum  Science  and  Technology, Universit\"at  Stuttgart,  Pfaffenwaldring  57,  70569  Stuttgart,  Germany}

\author{Tim Langen}
\email{tim.langen@tuwien.ac.at}

\affiliation{5. Physikalisches  Institut  and  Center  for  Integrated  Quantum  Science  and  Technology, Universit\"at  Stuttgart,  Pfaffenwaldring  57,  70569  Stuttgart,  Germany}

\affiliation{Vienna Center for Quantum Science and Technology, Atominstitut, TU Wien,  Stadionallee 2,  A-1020 Vienna,  Austria}

\begin{abstract}
An important effort is currently underway to extend optical cycling and laser cooling to more molecular species. Significant challenges arise in particular when multiple nuclear spins give rise to complex, resolved hyperfine spectra, as is the case for several molecular species relevant to precision tests of fundamental symmetries. We provide a detailed introduction to the use of optical spectra generated via serrodyne waveforms to address this complexity. We discuss our experimental implementation of these serrodynes, characterize their properties, and outline procedures to find optimized sideband configurations that generate strong laser cooling forces. We demonstrate the application of these techniques to barium monofluoride molecules and explore their prospects for the cooling of other species relevant to the study of fundamental physics.
\end{abstract}

\maketitle

\section{Motivation}
Laser cooling of molecules has recently made remarkable progress, and a large variety of molecules, from diatomics to polyatomics, can now be cooled~\cite{Fitch2021,Langen2023}. The central ingredient for this are diagonal Franck-Condon factors, which limit vibrational branching~\cite{DiRosa2004}. This renders optical cycling of thousands of photons possible using only a limited number of vibrational repumping lasers. Similarly, rotational branching can be limited by selection rules through an appropriate choice of molecular rotational levels involved in the optical cycle~\cite{Stuhl2008}.

However, the complexity of molecular structure extends well beyond vibrational and rotational branching, with hyperfine levels typically being addressed by laser sidebands to prevent population from accumulating in unaddressed dark states. These sidebands are typically created by a combination of acousto- and electro-optic modulators. This results in complex optical setups, where a significant fraction of the initial laser power is lost before it is applied to the molecules. These setups are also severely limited in terms of the spectra that can be generated, e.g., for optimized cooling and molasses techniques~\cite{Rockenhaeuser2024, Cheuk2018,Truppe2017}. 
Addressing all levels becomes particularly challenging for molecules with multiple nuclear spins~\cite{Kogel2021}, where the complexity of a resolved hyperfine structure may prevent optical cycling altogether. The latter is particularly relevant for many molecular species with potential for testing fundamental symmetries~\cite{Demille2008,Flambaum2014,Udrescu2021,Grasdijk2021,Zeng2023,ArrowsmithKron2024,Gaul2024}. 

\begin{figure}[tb]
    \centering
    \includegraphics[width=0.99\columnwidth]{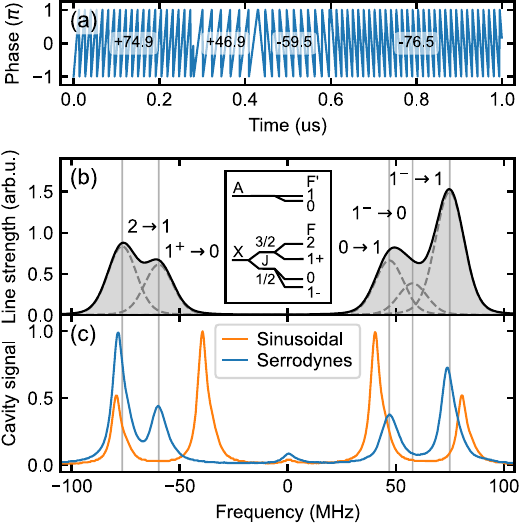}
    \caption{(a) Example of a serrodyne waveform that is imprinted onto laser light using a fiber-coupled EOM. Each linear phase shift corresponds to a frequency shift of the laser from its carrier frequency that is proportional to the slope of the waveform. By concatenating several segments with different slopes, time-sequenced spectra can be realized, with the amplitude of the individual frequencies set by the duration of the corresponding segments. (b) Hyperfine spectrum of the quasi-closed \gs$(N=1)\rightarrow$ \exs$(J'=1/2^+)$\, cycling transition in BaF. Here, $N$ and $J'$ are rotational angular momentum quantum numbers. The inset shows the corresponding level structure, with $J=N\pm1/2$ an intermediate angular momentum in the \gs state, and $F$ and $F'$ denoting the hyperfine angular momentum in the \gs and \exs\, states, respectively~\cite{Rockenhaeuser2023}. For efficient optical cycling, laser frequencies must be precisely matched to the resulting transitions.  (c) Optical spectra created using a sinusoidal drive of the EOM (orange line), and through the serrodyne waveform (blue line) from (a). While the restriction in terms of amplitudes and frequency offsets for the former lead to a significant mismatch, the latter closely matches the transition frequencies of the spectra in (b).}
    \label{fig:serrodyneprincipleandcycling}
\end{figure}

Recently, we have demonstrated that serrodynes are a powerful and easy-to-implement alternative to these setups for laser cooling~\cite{Rockenhaeuser2024}. Such serrodynes produce time-sequenced optical spectra that can be precisely tuned to the hyperfine structure of any typical molecular species. In doing so, they contribute to the effort to address increasingly complex molecules while also leading to a significant simplification of optical setups and greater flexibility for the optimization of the generated spectra for less complex species. In this work, we discuss our implementation of serrodynes in detail, present measurements using barium monofluoride (BaF) molecules highlighting prospects and limitations of serrodynes, and introduce procedures to find optimized sideband configurations that deviate from a given molecular spectrum to further enhance the laser cooling forces. Our results experimentally verify previous theoretical observations~\cite{Kogel2021,Tarbutt2015,Devlin2018,Alauze2021}, in which simulations favor the use of fewer, carefully tuned, laser sidebands to achieve stronger laser forces in molecules with complex hyperfine spectra.   

\section{Serrodynes}
To realize sustained optical cycling, it is important to address all hyperfine levels in the ground state of a molecule by a combination of optical drive~\cite{Fitch2021} and state mixing techniques~\cite{Yeo2015}. If this is not achieved, a significant population can accumulate in dark hyperfine levels and cycling stops before a sufficient number of photons can be scattered. The usual approach to address the optics side of this complication is to modulate laser light with a combination of acousto-optic modulators (AOMs) and sinusoidally driven electro-optic phase modulators (EOMs). This produces permanent sidebands with frequency offsets that are spaced by the modulation frequency, with amplitudes related to the power of the respective driving frequencies~\cite{SalehTeich}. 

Although powerful and widely used, this approach has several important limitations. First, it is limited to very specific spectra because, for example, the amplitudes of the sidebands generated by a sinusoidally driven EOM are given by Bessel functions and cannot be freely chosen to match the spectrum of a given molecule. Moreover, particularly for strong modulation, a considerable part of the total laser power can be lost to unwanted higher-order sidebands, which may even unintentionally drive undesired molecular transitions. Second, the modulation is typically realized using free space optics, which requires complex, bulky, and potentially unstable setups. These lead to a further loss of laser power due to imperfect coupling efficiencies of the various optical elements. 

The alternative approach that we use here is to drive EOMs using serrodyne waveforms. Such serrodynes have been widely used to tailor electromagnetic spectra~\cite{Cumming1957, Johnson2010,Balla2023}, including for the manipulation of molecules~\cite{Rogers2011}. The approach is based on the observation that 
an applied phase shift $\phi(t)=\alpha\, t$ that changes linearly with time, leads to a modulated light wave with  $E(t)=E_0 \exp{i(\omega_0 + \alpha)t}$. Here, $E_0$ is the amplitude of the electric field and $\omega_0$ its original frequency. Hence, $\omega_0$ is shifted according to the slope $\alpha$ of the phase shift. 

The linear phase increase required for this frequency shift can be realized by applying a voltage ramp to an EOM. Since this voltage cannot increase indefinitely, a periodic sawtooth voltage with the required slope is used instead to generate the desired frequency shift. For a target frequency offset of $\alpha$, the phase ramps are designed with an amplitude of $n\times[-\pi,\pi]$ and a period of $2\pi/n\alpha$, where $n$ is an integer. This setup allows the phase to periodically reset with each cycle while maintaining a stable linear frequency shift~\cite{Rogers2011}.

By concatenating several waveform segments with different slopes, nearly arbitrary optical spectra can be created. An example of this principle is shown in Fig.~\ref{fig:serrodyneprincipleandcycling}. It is important to observe that this approach realizes time-sequenced spectra, where only one frequency in the spectrum is present at any given time. However, if the switching between these frequency components is fast enough, the response of the molecules is only weakly affected compared to permanently present sidebands. We will discuss below how optical cycling and laser cooling respond to this switching rate. 

A related approach has recently been realized by driving EOMs with completely arbitrary waveforms generated by the Gerchberg-Saxton algorithm~\cite{Holland2021}. However, since serrodynes can potentially be implemented without any arbitrary waveform generator using only nonlinear elements, easily realizing shifts of many GHz~\cite{Houtz2009,Lauermann2016}, we consider them more accessible and versatile for the use in many experiments. Moreover, as we will show below, imperfections that arise in their implementation can conveniently be exploited to enhance the resulting laser cooling forces. 

The common advantage of both of these methods, serrodynes and Gerchberg-Saxton waveforms, is that optical setups can be fully integrated. For example, a fiber-coupled seed laser can directly feed into a fiber-coupled EOM, which is subsequently amplified in another fiber-coupled laser system and, if required, even frequency-doubled, before being delivered to the experiment. This allows for very compact, stable, and power-efficient optical setups, which is particularly important for the laser cooling of molecules, where dozens of high-power cooling and repumping lasers are sometimes required in parallel. 

\section{Implementation}
\begin{figure}[tb]
    \centering
    \includegraphics[width=1\columnwidth]{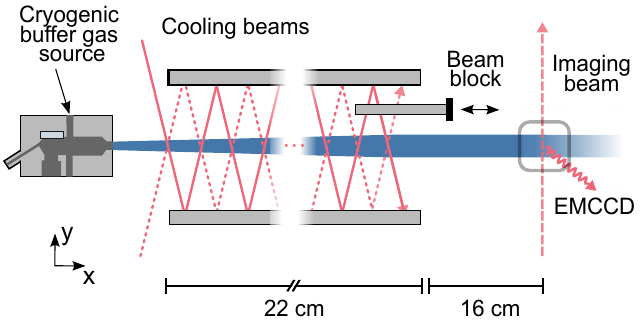}
    \caption{Experimental setup for optical cycling and laser cooling of BaF molecules. A molecular beam is created using a cryogenic buffer gas cell and travels along the x-direction. During magnetically-assisted laser cooling, it interacts with two counter-propagating laser beams in the transverse y-direction that are retro-reflected multiple times over a distance of up to $22\,$cm. To observe only optical cycling and inhibit laser cooling, one of the two counter-propagating laser beams (dotted) can be removed. The length of the interaction region is controlled using a movable beam block, ranging from a single laser beam pass up to the full $22\,$cm. After optical cycling or transverse cooling in this interaction region, the molecular beam expands for a further $16\,$cm before being imaged by recording the fluorescence induced by an imaging laser beam with an EMCCD camera.}
    \label{fig:setup}
\end{figure}

In the following, we demonstrate the use of serrodynes for optical cycling and laser cooling, using \baff{138} molecules as a concrete example. This species has a relatively high mass, which is of great interest for various types of precision measurement applications~\cite{Aggarwal2018, Altuntas2018, Vutha2018,Flambaum2014,Gaul2024}. However, this high mass and a resolved hyperfine splitting in the excited state also render the molecule challenging to cool, requiring careful optimization of the laser cooling forces. As we demonstrated in previous work~\cite{Rockenhaeuser2024,Kogel2024}, this optimization can be conveniently achieved using serrodynes. 

Similar to other alkaline-earth monofluorides, the even isotopologues of BaF feature four hyperfine levels in the ground state and two in the excited state, which are, however, all optically resolved~\cite{Rockenhaeuser2023,Denis2022,Bu2022}. Taken together, for a single vibrational state, this leads to the 12+4 level scheme depicted in Fig.~\ref{fig:serrodyneprincipleandcycling}b, with 12 ground state magnetic sublevels and 4 excited state magnetic sublevels per vibrational manifold. Of the six possible transitions between hyperfine levels, one is nearly forbidden~\cite{Rockenhaeuser2023}, leading to a total of five relevant transitions. One of these transitions overlaps with two stronger ones and is therefore always driven off-resonantly when the other two are addressed with typical powers. Conventionally, and again similar to other alkaline-earth monofluorides, a total number of four laser sidebands resonantly addressing the resulting hyperfine spectrum in Fig.~\ref{fig:serrodyneprincipleandcycling}b is thus required to avoid dark states. However, these four transitions are far from being equally spaced and are therefore difficult to address with a sinusoidally driven EOM alone (Fig.~\ref{fig:serrodyneprincipleandcycling}c), requiring an additional AOM in conventional schemes~\cite{Zeng2024}, or the use of serrodynes. Moreover, as detailed below, to optimize laser cooling it may be required to realize fully tunable configurations with only three or fewer sidebands with controllable amplitudes, for which serrodynes are the only viable option. 

Our optical setup to realize the required laser frequencies and their sidebands driving the quasi-closed \gs$(\nu, N=1)\rightarrow$ \exs$(\nu', J'=1/2^+)$ transition between various vibrational states $\nu$ and $\nu'$ in BaF~\cite{Albrecht2020}, consists of external cavity diode lasers in the near-infrared spectral region, which are amplified using tapered amplifiers (TAs). Here,   $N$ and $J'$ are rotational angular momentum quantum numbers in the ground and excited states, respectively. 

To generate sidebands for cycling and cooling ($\nu=0\rightarrow\nu'=0$), we fiber-integrate the entire optical setup as described above. The fiber-coupled EOM used (iXblue NIR-MPX800-LN-05) is very power sensitive and is thus placed between the diode laser and the TA. Is has an attenuation of $4\,$ dB, producing an output power of $10$,mW, which is sufficient to seed the TA. We do not observe any non-linear effects in the TA resulting from the phase-modulated seed laser (see the appendix). The sidebands for repumping ($\nu\rightarrow\nu'=\nu-1$) and depumping transitions ($\nu=0\rightarrow\nu'=1$), as well as for detection via the cooling transition ($\nu=0\rightarrow\nu'=0$) are generated by free-space EOMs (QUBIG PM8-NIR). These are driven sinusoidally, have a much higher laser damage threshold than fiber-coupled EOMs, and are thus placed after the respective TAs. 

The rest of the experimental setup used for realizing optical cycling and laser cooling is summarized in Fig.~\ref{fig:setup} and further detailed in previous work~\cite{Albrecht2020, Rockenhaeuser2024,Kogel2024}. In short, a cold molecular beam of BaF is generated in a cryogenic buffer gas cell at $3.5\,$K. The beam travels through an interaction region that is up to $22\,$cm long, with retro-reflected laser light in the transverse direction. 
For Sisyphus-type magnetically assisted laser cooling, two counterpropagating laser beams are used. In contrast to this, during measurements of optical cycling, laser cooling forces are inhibited by blocking one of these counter-propagating beams to remove the intensity standing wave that facilitates such cooling.  The beam diameters are $3.1\,$ mm full width at half maximum, and each beam undergoes up to $17$ round-trip passes, with the number of round-trip passes --- and thus the overall length of the cooling region ---  controlled by a movable beam block. Further downstream, the transverse profile of the molecular beam can be imaged after $16\,$cm of free expansion in a detection region via laser-induced fluorescence captured onto an EMCCD camera~\cite{Rockenhaeuser2023}. 

\begin{figure}[tb]
    \centering
    \includegraphics[width=\columnwidth]{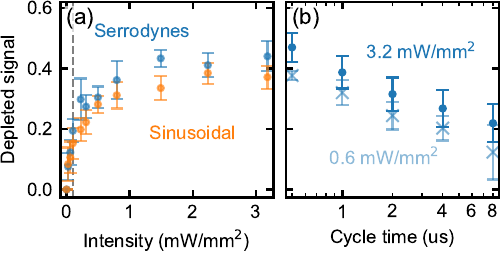}
    \caption{Optical cycling characterized using depletion of the \gs\,$(\nu=0)$ state population. Cycling and detection are realized with two independent cooling lasers ($\nu=0\rightarrow\nu'=0$). The depletion of the signal is directly proportional to the number of scattered photons.   
    (a)~Depleted signal as a function of laser power for both sinusoidal (orange) and serrodyne (blue) modulation of the laser beam. For the given laser beam parameters, the saturation intensity of the equivalent two-level system corresponds to a power of $I_\mathrm{sat}=5.8\,\mu$W/mm$^2$, which for the given $12+4$ level system translates to $I_{\mathrm{sat,eff}}=105\,\mu$W/mm$^2$~\cite{Fitch2021} (dashed vertical line). For higher powers the depletion starts to saturate, as expected. Both sinusoidal and serrodyne modulation perform the same within error bars, despite the time-sequenced nature of the serrodynes. (b)~Depleted signal as a function of the serrodyne cycle time. Shorter cycle time corresponds to a higher repetition rate of the individual serrodyne segments. 
    Compared to the $56\,$ns excited state lifetime of the \exs\, state in BaF, the shortest cycle times examined are still considerably longer than this timescale. In line with this, we observe a monotonous increase of the number of scattered photons for decreasing cycle time. This increase does not saturate independent of the laser power used, indicating that even faster switching between segments could further increase the scattering rate.}
    \label{fig:opticalcyclingplots1}
\end{figure}

\section{Optical cycling}

In a first step, we study optical cycling of the molecules to demonstrate the capability to realize a quasi-closed transition using the serrodynes. 

To this end, we begin by using only a single pass of the cooling laser without any repumpers and generate a serrodyne spectrum that has four frequency components that match the four peaks in the hyperfine spectrum of BaF, as illustrated in Fig.~\ref{fig:serrodyneprincipleandcycling}c.
An iris is used to shape the beam into a uniform intensity distribution with a diameter of 2 mm. After cycling up to $15$ photons while passing through the single beam in the interaction region, the molecules are imaged onto the EMCCD camera in the detection region, using light resonant to the same cooling transition. The detection
thus probes the population in $\nu=0$. Leakage into higher vibrational ground states with $\nu\geq1$ due to optical cycling can be observed as a depletion of the corresponding signal, which is directly proportional to the number of scattered photons. 

Fig.~\ref{fig:opticalcyclingplots1} illustrates the depleted signal recorded on the EMCCD camera as a function of various parameters. In Fig~\ref{fig:opticalcyclingplots1}a, we find that both serrodyne modulation and sinusoidal modulation perform similarly within error bars as a function of laser power, despite the time-sequenced nature of the serrodynes. This is due to the comparably low powers around saturation studied here, which limit power broadening of the spectral lines, and thus, improve the relative performance of the more resonant serrodynes. However, the limitation of their time-sequenced nature is evidenced in Fig.~\ref{fig:opticalcyclingplots1}b, where we observe a monotonous increase in the depleted signal with shorter cycle times of the serrodyne waveform. Here, shorter cycle times correspond to a higher repetition rate of the individual frequency components. Since the shortest cycle times studied are still longer than the timescale set by the inverse of the maximum scattering rate $R_\mathrm{max}=\Gamma/4=1/224\,$ns for the 12+4 level system, faster switching is expected to further improve the scattering rate and approach saturation only when these two timescales are comparable. Here, $\tau = 1/\Gamma = 56\,$ns is the lifetime of the excited state \exs, of BaF~\cite{Aggarwal2019}.

Next, we let the molecules interact with both the cooling laser and the first vibrational repumping laser throughout a varying fraction of the full interaction region, scattering up to $250\,$ photons. After that they are imaged with the combined depumping laser and the first vibrational repumping laser in a resonant-Raman optical cycling scheme~\cite{Rockenhaeuser2023,Shaw2021}.
This imaging scheme is effectively sensitive to the combined populations in both the $\nu=0$ and $\nu=1$ ground states. The resulting depletion signal thus again probes the number of scattered photons, this time through the leakage into vibrational states with $\nu\geq 2$ and a lower-lying electronic loss state~\cite{Hao2019}. 

\begin{figure}[tb]
    \centering
    \includegraphics[]{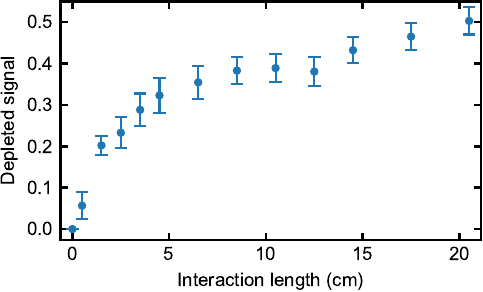}
    \caption{Sustained optical cycling characterized using depletion of the combined populations in the \gs\,$(\nu=0)$ and \gs\,$(\nu=1)$ states as a function of the length of the interaction region. More depletion is directly proportional to a higher number of scattered photons. For a constant scattering rate, this plot would scale linearly. However, due to the increasing depletion the scattering rate drops as the molecules travel through the interaction region. In addition, due to imperfect alignment and losses during each reflection, the laser beams in the interaction region gradually lose power, thereby leading to a decreasing scattering rate, and thus a smaller rate of increase for the depletion.}
\label{fig:opticalcyclingplots2}
\end{figure}

In Fig.~\ref{fig:opticalcyclingplots2}, we characterize optical cycling in this configuration as a function of the interaction length. Ideally, for a constant scattering rate, the depleted signal is expected to increase linearly. However, as
the population is pumped into the mentioned loss states, the population in the $\nu=0$ and $\nu'=0$ ground states, which are the ones dominating the cycling, gradually decreases and so does the corresponding scattering rate. Additionally, due to imperfect alignment and losses during each reflection of the lasers along the interaction region, about 60\% of the initial power is lost, thereby contributing to the reduction of the scattering rate, and thus also to the slower increase of the depletion~\cite{Rockenhaeuser2024}. 

As the repumping laser effectively contributes very little to the overall number of scattered photons, we find no difference between driving it with a serrodyne or a conventional sinusoidal waveform, as long as sufficient power is used. For simplicity, we thus use a sinusoidal drive for all repumping and depumping lasers in the following. 

\begin{figure*}[tb]
    \centering
    \includegraphics[width=0.98\textwidth]{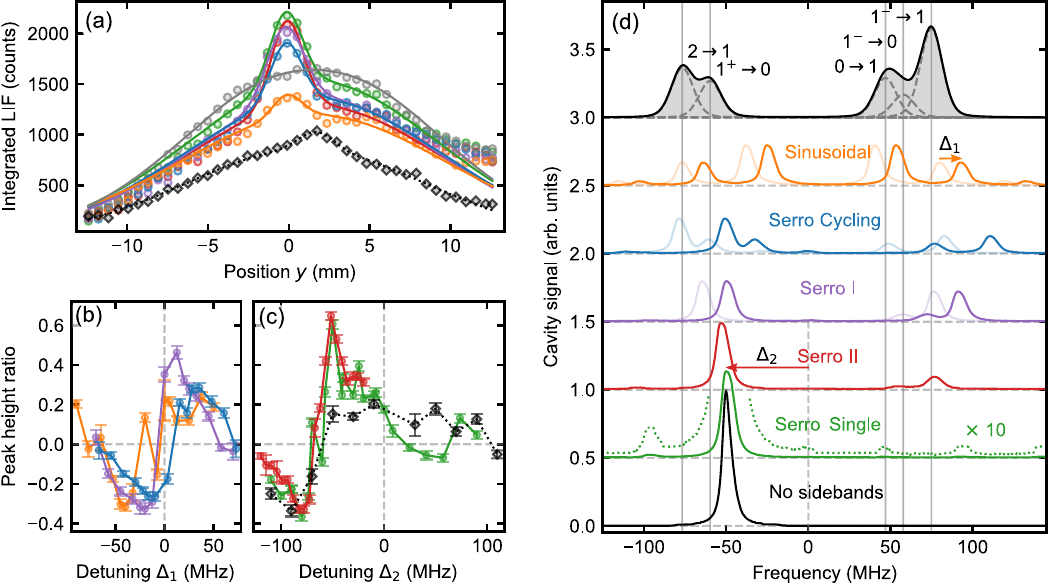}
    \caption{Characterization of laser cooling for different sideband configurations of the cooling laser. (a)~Integrated line profiles of the transverse cross-sections of the molecular beam obtained via fluorescence imaging~\cite{Rockenhaeuser2023}. The colors correspond to different sideband configurations visualized in (d). Gray circles represent the unperturbed molecular beam. The solid lines are fits to the data using a double-Gaussian function described in Ref.~\cite{Rockenhaeuser2024}. The data for each sideband configuration was recorded at the optimum detuning for cooling efficiency. The data for no sidebands (black diamonds) was collected on a separate day compared to the other measurements, resulting in a slightly different peak position caused by daily variations in the optical alignment. (b,c)~Cooling efficiency, inferred using the peak height ratio of the cooled and uncooled molecular distributions, as a function of laser detuning. Positive (negative) cooling efficiency denotes laser cooling (heating). Note that due to the different shapes of the individual optical spectra, the points of vanishing force are located at different absolute frequencies. Serro II demonstrates the best performance overall but results in slightly higher losses with the current laser arrangement, which causes it to be outperformed by Serro Single in (a). Detunings $\Delta_1$ and $\Delta_2$ correspond to cases where the entire spectrum or only a single sideband are shifted, respectively. (d)~Optical spectra of the sideband configurations, measured using a Fabry-Perot cavity, compared to the simulated hyperfine spectrum of \baff{138} with Doppler broadening (top). 
    Each configuration is represented using the optimal detuning for cooling efficiency obtained from (b,c), and used in (a), and illustrates the exact positions of the laser frequency components with respect to the hyperfine transitions (vertical gray lines). To scan the laser detuning, either the entire optical spectrum is shifted by $\Delta_1$ in frequency across the hyperfine spectrum (Sinosoidal, Serro Cycling and Serro I), or only the dominant frequency component is detuned by $\Delta_2$, while the components addressing $F=0 \rightarrow F'=1$ and $F=1^-\rightarrow F'=1$ transitions remain effectively unchanged.}
    \label{fig:cooling-diff-sidebands}
\end{figure*}

\section{Transversal Sisyphus laser cooling}

In our previous work~\cite{Rockenhaeuser2024}, we presented results for optimized magnetically assisted laser cooling forces using serrodyne-modulated optical spectra. Here, we describe the optimization procedure, observations, and principles involved that are generalizable to other species, in particular those with complex hyperfine structure. 

For transverse laser cooling, we apply cooling and both repumping lasers throughout the full interaction region, with powers that greatly exceed saturation. Counterpropagating pairs of beams are realized to form the characteristic intensity standing wave required for magnetically-assisted laser cooling.

Fig.~\ref{fig:cooling-diff-sidebands} characterizes the resulting laser forces for a variety of sideband configurations to illustrate and explain the optimization procedure. The cooling efficiency is inferred by fitting a double Gaussian function to the cooled molecular peak and the uncooled background envelope and then calculating the ratio of peak heights~\cite{Rockenhaeuser2024}. Simulations suggest that this peak height ratio scales approximately linearly with the average force applied to the molecular beam in our setup, although the exact details will depend sensitively on the specific experimental implementation and alignment.

Conventional sinusoidally modulated sidebands, illustrated in orange, result in the lowest cooling efficiency among all configurations studied. Their behavior as a function of laser detuning in Fig.~\ref{fig:cooling-diff-sidebands}b deviates from the characteristic behavior expected from Sisyphus-type forces and instead oscillates irregularly between heating and cooling. We attribute this to a competition between the weak second-order components and strong first-order components of the sinusoidal drive, which, for certain detunings, may be oppositely detuned (red and blue) to different states. For instance, for small positive detunings $\Delta_1$ above $0\,$MHz, the $-1\,$st order sideband is blue detuned to the $F=1^+ \rightarrow F'=0$ transition leading to cooling, while the $-2\,$nd order sideband is simultaneously red detuned to it, leading to heating. While such competition always occurs in the presence of multiple sidebands, the intuitive improvement upon this is to mimic the hyperfine spectrum of the molecule in the applied optical spectrum using serrodynes, as illustrated in blue (Serro Cycling). This leads to considerably higher forces as observed in the height of the corresponding cooled molecular fraction in Fig.~\ref{fig:cooling-diff-sidebands}a. In fact, this optical spectrum is the simplest optimization that can be realized and should be easily applicable to most molecules, including those with highly complex spectra. Its behavior as a function of detuning now also follows the expected trend, without artifacts of competition between heating and cooling due to different frequency components. 

\begin{figure}
    \centering
    \includegraphics[]{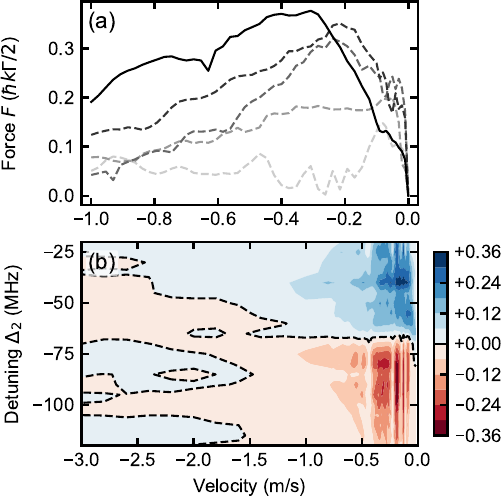}
    \caption{Simulation and optimization of Sisyphus-type cooling forces in \baff{138} molecules in units of the maximum force $\hbar k\Gamma/2$ of an equivalent two-level system. (a) Typical force traces in the optimization process that are integrated and maximized in the shown velocity range by varying parameters such as power, magnetic field and sideband configuration. The dashed lines correspond to typical iterative optimization steps leading to the  the final result (solid line) referred to as Serro I (see Fig. \ref{fig:cooling-diff-sidebands}). 
    (b) Simulated force profile for the Serro II sideband configuration with the detuning $\Delta_2$ of the left-most sideband. The color scale indicates the magnitude of the force, and the dashed line shows where the force changes its sign (see Fig.~\ref{fig:cooling-diff-sidebands}c).}.
    \label{fig:simulation}
\end{figure}

For further optimization of the forces, we perform simulations based on the optical Bloch equations~\cite{Kogel2021,Rockenhaeuser2024}. To this end, force profiles such as those shown in Fig.~\ref{fig:simulation} are calculated for all relevant parameters. By optimizing for the largest force in a desired velocity range, favorable conditions for laser cooling can be identified.

In particular, informed by our simulations in Ref.~\cite{Kogel2021}, and previous pioneering work~\cite{Tarbutt2015,Devlin2018,Alauze2021}, we explore the further increase of the cooling force using only three, rather than the previous four sidebands. An optimized configuration for these three sidebands can be deduced from simulations and is illustrated in Fig.~\ref{fig:cooling-diff-sidebands} in purple (Serro I). For this configuration, from Figs.~\ref{fig:cooling-diff-sidebands}a and b, we observe a further improvement in cooling efficiency. Comparing the optical spectra and detuning behavior for Serro I and Serro Cycling, we infer that the dominant forces are produced by the sideband addressing the $J=3/2$ ground-state manifold (which includes the $F=2 \rightarrow F'=1$ and $F=1^+ \rightarrow F'=0$ transitions). Here, $J=N\pm1/2$ is an intermediate angular momentum in the ground state (see Fig.~\ref{fig:serrodyneprincipleandcycling}b). Thus, to reduce competition between sidebands, to concentrate the available laser power in this frequency component, and to still address the other $J=1/2$ ground state manifold to avoid dark states, we implement a second configuration with three components. This is illustrated in red (Serro II), with two weak frequency components resonantly repumping the $J=1/2$ manifold, while only the strong third component is scanned over the $J=3/2$ manifold to generate and tune the forces. This scheme yields the highest cooling efficiency. 

\begin{figure*}[tb]
    \centering
    \includegraphics[width=0.98\textwidth]{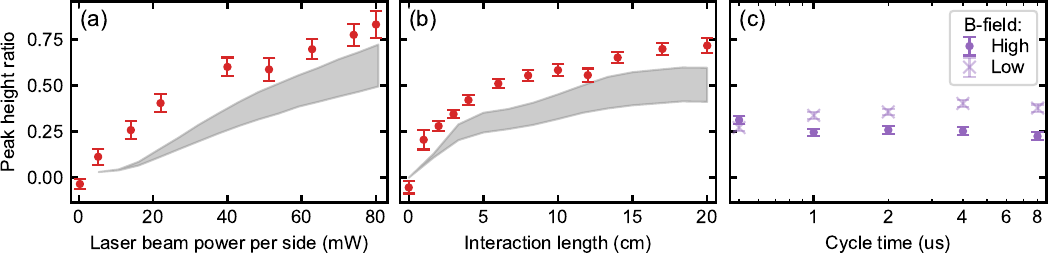}
    \caption{Laser cooling efficiency, as characterized by the peak height ratio. The dots are experimental data while the shaded regions are simulation results. (a)~Cooling efficiency as a function of laser power. The efficiency increases monotonically with laser power and does not approach any saturation. (b)~Cooling efficiency as a function of the length of the interaction region. The cooling efficiency quickly rises for the first $5\,$cm, and only changes very little for longer interaction lengths, due to the decreasing power of the reflected laser beams. (c)~Cooling efficiency as a function of the cycle time of the serrodyne modulating signal for low ($1\,$G) and high ($5\,$G) magnetic fields. In contrast to optical cycling, hardly any change in cooling efficiency is observed as a function of the cycle time.}
    \label{fig:coolingcharacterization}
\end{figure*}

To further investigate the mechanisms active for Serro II, and to highlight how fewer frequency components yield stronger cooling forces due to less competition between different transitions, we next use a single serrodyned frequency component, shown in green, which scans over the entire hyperfine spectrum. The dotted green line in Fig~\ref{fig:cooling-diff-sidebands}c shows a magnified view of the same optical spectrum, where in addition to the single serrodyned component, we observe several weak higher-order contributions due to the limited bandwidth of our arbitrary waveform generator (Tektronix AWG 2041) and amplifier (Minicircuits ZHL-3A+) used to generate the serrodyne modulation. These higher-order contributions can in principle be reduced using higher-bandwidth devices (see the appendix) and a more careful calibration of the frequency response of the experimental setup~\cite{Holland2021}. However, we observe that, despite being very weak and off-resonant, the high-order contributions can induce sufficient repumping of the non-essential transitions of the $J=1/2$ manifold, resulting in similar forces as observed for Serro II. Due to the lower scattering rate, Serro Single even leads to fewer losses and thus slightly outperforms Serro II in absolute signal strength. We do, however, expect this advantage to disappear when these losses are compensated by an additional repumping laser. Furthermore, in Fig.~\ref{fig:cooling-diff-sidebands}c, as we scan the single serrodyne over the entire hyperfine spectrum, we observe two distinct heating and cooling profiles, respectively. The dominant one centered on $\Delta_2=-66\,$ MHz corresponds to the $J=3/2$ manifold, while a second weak profile centered on $\Delta_2=63\,$ MHz corresponds to $J=1/2$. This behavior clearly illustrates the important role of the $F=2\rightarrow F'=1$ and $F=1^+ \rightarrow F'=0$ transitions in realizing the majority of the Sisyphus forces in \baff{138}, which is consistent with our simulations. 

In general, the ideal situation arises for blue detuning of the relevant sideband from both these $J=3/2$ transitions. This can be understood from the observation that $F=2\rightarrow F'=1$ has a stronger transition amplitude, and thus a larger detuning is required to maximize Sisyphus forces arising from this transition. As this transition addresses the highest ground-state multiplicity, it contributes the strongest to the forces. Placing the sideband next to $F=1^+ \rightarrow F'=0$ thus leads both to cooling forces from this transition and, at the same time, to the largest possible cooling from $F=2\rightarrow F'=1$. An additional fourth sideband, as discussed in some of the conventional sideband configurations above, would either be red-detuned to some of the transitions and would, hence, decrease the forces again, or be blue-detuned with a larger detuning, contributing less effectively to the overall force. Additionally, a higher number of sidebands would lead to enhanced coherent fluctuations in the dynamics of the intermolecular population that also reduce the magnitude of the force. This behavior is expected to be reproduced in all bosonic earth-alkali monofluoride molecules, such that the principles observed here can be readily transferred to other species.  

Finally, to further demonstrate the role of the higher-order frequency components in closing the optical cycle, we scan the bare unmodulated laser over the hyperfine spectrum. This is illustrated in black in Figs.~\ref{fig:cooling-diff-sidebands}a and c. Surprisingly, we still observe sizable laser forces, despite the fact that only very few photons can be scattered on average, before molecules are lost into unaddressed hyperfine dark states. While this observation highlights the strength of the Sisyphus forces realized in BaF, the unmodulated laser produces considerably less cooling efficiency compared to the single serrodyne. This verifies that the higher-order peaks in the optical spectrum are indeed crucial for repumping the non-essential transitions that do not significantly contribute to the laser forces.

Interestingly, Figs.~\ref{fig:cooling-diff-sidebands}b and c also reveal an asymmetry with stronger cooling compared to heating in general. Although heating remains stable within the error bars for different sideband configurations, the cooling forces vary significantly between configurations. We attribute this to two sources -- one, the simulations used to deduce the optimum configurations for different sidebands were optimized exclusively for cooling forces. Two, Sisyphus heating in molecules requires red-detuned light. Since most of the force is generated by the $J=3/2$ manifold, the corresponding components need to be red-detuned. However, in this situation, the components addressing the $J=1/2$ levels are still blue-detuned to the former, leading to competition. In contrast, during cooling, all frequency components can be blue-detuned simultaneously to the $J=3/2$ manifold, leading to less competition and thus stronger forces. 

Finally, using Serro II as the optimum sideband configuration, Fig.~\ref{fig:coolingcharacterization} characterizes the cooling efficiency as a function of the laser power, the length of the interaction region, and the cycle time of the serrodyne modulation. We observe a monotonically increasing cooling efficiency with increasing laser power, suggesting that the forces are not yet reaching saturation, and more laser power may achieve even higher fractions of cold molecules. However, with increasing length of the interaction region, the cooling efficiency approaches saturation, which may again be attributed to imperfect alignment and the decreasing power along the interaction region due to reflection and transmission related losses. Notably, we do not observe any sizable dependence of the cooling efficiency on the serrodyne cycle time, irrespective of the magnetic field strength, verifying that the Sisyphus forces realized here --- in contrast to optical cycling or Doppler forces --- do not strongly depend on the photon scattering rate.

\section{Conclusion}

We have discussed how serrodynes can be a robust and useful tool for addressing the hyperfine structure in molecular laser cooling experiments. Their fiber-based implementation enables simplified laser setups, and can readily be incorporated into existing experiments without the need for major infrastructural changes. While our measurements of optical cycling show that they achieve photon scattering rates that are competitive with conventional sinusoidal modulation schemes, their time-sequenced nature leaves room for significant further improvements beyond these conventional methods with the use of faster waveform generators. This will be particularly important when using them to realize and optimize Doppler forces, e.g. for magneto-optical trapping. Similar principles can be applied to other laser cooling tasks, for example, for complex ions~\cite{Fan2019,Notzold2022,Huang2024}.

The flexibility to generate arbitrary optical spectra has also allowed us to engineer strong Sisyphus forces for BaF molecules, and in the process verify important principles that will inform the optimization of such forces for other molecular species. In line with our simulations in Ref.~\cite{Kogel2021}, our measurements verify that fewer frequency components dedicated to addressing only the strongest transitions can lead to higher cooling efficiencies compared to exactly replicating the molecular spectra with the laser sidebands, predominantly by mitigating competition between heating and cooling due to different frequency components. In particular, we note that the majority of laser forces are generated by the $J=3/2$ manifold of the hyperfine spectrum in \baff{138}. To optimize the forces, the laser power needs to be concentrated in the frequency component addressing these transitions, whereas the remaining transitions only need to be weakly repumped to sufficiently close the optical cycle. 

This result is particularly important for molecules with complex hyperfine spectra, for instance, the odd isotopologues of BaF, where the number of hyperfine transitions can be prohibitively large to address them individually. In such systems, applying the principles developed here and in Ref.~\cite{Kogel2021}, optimization of the laser forces would require us to identify the ground states with the highest multiplicities (or, equivalently, states that contribute most to the forces) and address them strongly with dedicated frequency components, while weakly repumping the rest. In the context of the odd isotopologues, state preparation for precision measurements of nuclear spin-dependent parity violation, as well as efficient cycling readout for the same~\cite{Altuntas2018,AltuntasPRA,Lasner2018}, are other challenging task that may be simplified using tailored optical spectra.

This flexibility offered by serrodynes is also a crucial topic as different molasses techniques, for instance, allow better detection of molecules~\cite{Cheuk2018}, while others may allow for more efficient cooling below the Doppler limit~\cite{Truppe2017}. Similarly, deep laser cooling of trapped molecules requires the implementation of blue-detuned MOT schemes with intermittent $\Lambda$-enhanced molasses~\cite{Burau2023,Li2024}. In such scenarios, the optical spectra need to switch quickly between different configurations, which can be readily achieved using serrodyne modulation. 

\section*{Acknowledgments}
We are indebted to Tilman Pfau for generous support, thank Einius Pultinevicius for contributions in the early stage of the experiment. We acknowledge discussions with Olivier Grasdijk, Mangesh Bhattarai, and Connor M. Holland about techniques to synthesize optical spectra. This project has received funding from the European Research Council (ERC) under the European Union Horizon 2020 research and innovation program (Grant agreement No. 949431), Vector Stiftung, Carl Zeiss Stiftung, the RiSC initiative of the Ministry of Science, Research and Arts Baden-W\"urttemberg, and was funded in whole or in part by the Austrian Science Fund (FWF) 10.55776/PAT8306623.

\bibliography{biblio}

\section*{Appendix}

\begin{figure}[b]
    \centering
\includegraphics[width=0.99\columnwidth]{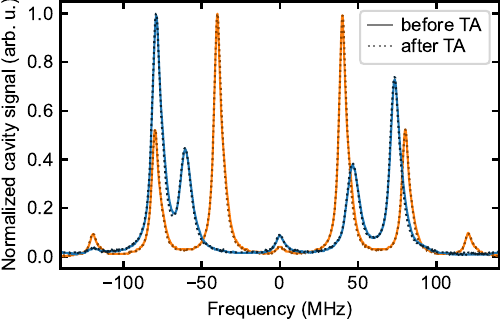}
    \caption{Comparison of the optical spectra before (solid line) and after (dotted line) the tapered amplifier (TA) for the cycling serrodynes (blue) and sinusoidal (orange) sidebands. As there is no difference visible, the TA can safely be seeded with phase-modulated light without being damaged or resulting in non-linear effects.\vspace{-14pt}}
    \label{fig:TAcomparison}
\end{figure}

\subsection{Modulated injection into a tapered amplifier}

The flexibility to generate arbitrary spectra is only possible for the combination of a fiber-coupled EOM seeded into a TA, as free-space AOMs/EOMs are typically only designed to achieve sufficient modulation depths for specific frequencies. For the TAs to not be damaged by the seed, the light must not be intensity modulated, but can be phase-modulated~\cite{Zappala2014}. However, significant nonlinear processes can still occur. We have therefore examined the spectra before and after a TA using a Fabry-Perot cavity, as shown in Fig.~\ref{fig:TAcomparison}. Apart from higher-order harmonics, which arise from the finite bandwidth of the arbitrary wave form generator and amplifiers used during modulation of the fiber EOM (see next section), we do not observe any additional nonlinear processes in our setup. This analysis was conducted under various conditions, including no modulation, sinusoidal modulation, and sawtooth modulation with different numbers of sidebands and modulation depths. In all cases, no further nonlinear processes were detected. 

\begin{figure}[tb]
    \centering
    \includegraphics[width=0.98\columnwidth]{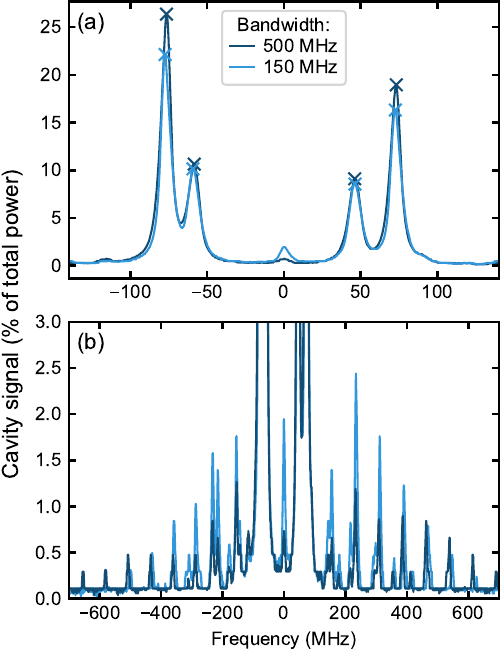}
    \caption{(a) Comparison of optical spectra (Serro Cycling) for two amplifiers with different bandwidths, as observed on a Fabry-Perot cavity. Higher bandwidth of the amplifier directly results in a higher ratio of the total power being distributed into the four main peaks instead into multiple higher-order frequency components. (b) Zoomed view of higher orders over a wider frequency range.}
    \label{fig:AmplifierComparison}
\end{figure}

\subsection{Higher-order components in the optical spectrum}
In our previous work~\cite{Rockenhaeuser2024}, we discussed the impact of the finite bandwidth of the AWG used to generate the modulating signal on the optical spectrum. The limited temporal resolution leads to a comb of weak higher-order frequency components, as observed in Fig.~\ref{fig:AmplifierComparison}b. In addition, the bandwidth of the amplifier that amplifies the serrodyne waveform can also lead to these. To exemplify this, we measured the optical spectra for different amplifiers with 500\,MHz (Minicircuits ZHL-1-2W+) and 150\,MHz (Minicircuits ZHL-3A+) bandwidths, shown in Fig.~\ref{fig:AmplifierComparison}, and found that, as intuitively expected, higher bandwidths can lead to a reduction of higher order components if this is desired. Alternatively, higher-order sidebands can be efficiently minimized by taking into account the transfer function of the waveform system~\cite{Holland2021}.

\subsection{Sideband configurations}
The data in Tab.~\ref{tab:serrodynedata} outline the sideband configurations as shown in Fig.~\ref{fig:cooling-diff-sidebands}, investigated at their optimal detunings (relative to the spectrum in Fig.~\ref{fig:cooling-diff-sidebands}d), along with the relative amplitudes of the individual sidebands and the corresponding maximum peak height ratios phr$_\text{max}$. Note that the actual power to which the molecules are exposed for each sideband will depend strongly on the characteristics and response of the frequency-generating setup, as discussed in more detail in the previous section.

\begin{table}[b]
    \centering
    \begin{tabular}{l| cccc | cccc | c}
    \toprule
    Type & \multicolumn{4}{c|}{Frequencies (MHz)}& \multicolumn{4}{c|}{Amplitudes (\%)}  &  phr$_\text{max}$ \\ \midrule
    Sinusoidal & 92 & 52 & -26 & -66 & 18 & 32 & 32 & 18  & 0.24\\
    Serro Cycling & 103 & 75 & -32 & -48 & 28 & 15 & 17 & 40 &  0.29\\
    Serro I & 89 & 68 & -48 &  & 43 & 9 & 47 &  &  0.46\\
    Serro II & 74 & 53 & -52 &  & 18 & 4 & 78 &  &  0.65\\
    Serro Single & -47 &  &  &  & 100 &  &  &  &  0.61\\
    No sidebands & -50 &  &  &  & 100 &  &  &  & 0.15\\ \bottomrule
    \end{tabular}
    \caption{Summary of the optical spectra used in this work.}
    \label{tab:serrodynedata}
\end{table}

\end{document}